\documentclass[fleqn,10pt]{wlscirep}
\usepackage[utf8]{inputenc}
\usepackage[T1]{fontenc}

\pdfoutput=1

\usepackage{hyperref}
\usepackage{amsmath}
\usepackage{amssymb}
\usepackage{graphicx}
\usepackage{color}
\usepackage{simplewick}
\usepackage[version=3]{mhchem}
\usepackage{bbold}
\usepackage{wasysym}
\usepackage{bm}
\usepackage{caption}
\usepackage{subcaption}

\renewcommand{\vec}[1]{{\boldsymbol #1}}

\begin{document}

\title{Spin liquid mediated RKKY interaction}

\newcommand{\sta}{SUPA, School of Physics and Astronomy,
University of St Andrews, North Haugh, St Andrews, Fife KY16 9SS, United Kingdom}

\newcommand{\col}{Institut f{\"u}r Theoretische Physik,
Universit{\"a}t zu K{\"o}ln, D-50937 Cologne, Germany}

%
\author[1,*]{Henry F. Legg}
\author[2]{Bernd Braunecker}
\affil[1]{\col}
\affil[2]{\sta}

\affil[*]{hlegg@thp.uni-koeln.de}


\begin{abstract}
We propose an RKKY-type interaction that is mediated by a spin liquid. If a spin liquid exists such an interaction could leave a fingerprint by ordering underlying localised moments such as nuclear spins. This interaction has a unique phenomenology that is distinct from the RKKY interaction found in fermionic systems; most notably the lack of a Fermi surface and absence of the requirement for itinerant electrons, since most spin liquids are insulators. We demonstrate that the interaction is predominately shaped by the lattice symmetries of the underlying spin liquid. As a working example we investigate the possible ordering of nuclear spins that interact through an underlying lattice of the two-dimensional spin-1/2 kagome antiferromagnet (KHAF), although the treatment remains general and can be extended to other spin liquids and dimensions. We find that several different nuclear spin orderings minimise the RKKY-type energy induced by the KHAF but are unstable due to a zero-energy flat magnon band in linear spin-wave theory. Despite this we show that a small magnetic field is able to gap out this magnon spectrum resulting in an intricate nuclear magnetism.
\end{abstract}

\maketitle


\section*{Introduction}

Spin liquids are characterised by a suppression of magnetic ordering at finite temperatures.\cite{bal:2010} Magnetic order in a spin liquid can be prevented by frustration; that is to say that competing forces prevent all interactions from being simultaneously minimised. For small spins, $S$, quantum fluctuations can suppress order at all non-zero temperatures leading to a quantum spin liquid. Quantum spin liquids can be distinguished by their high degree of long-ranged entanglement.\cite{savary2016quantum}
In this paper we show how spin liquid dynamics can mediate an interaction between another species of local moments.
Such an interaction is akin to the Ruderman-Kittel-Kasuya-Yosida (RKKY) interaction found in fermionic systems,\cite{ruderman:1954,kasuya:1956,yosida:1957} but appears here in a system that is an electronic insulator.
For instance a local spin liquid excitation can be produced through an exchange coupling to a different species of magnetic moments. Due to the correlated nature of the spin liquid this excitation can spread far over the system before coupling to another magnetic moment of the different species elsewhere in the material and, as a result, producing an effective long range Heisenberg type interaction between these magnetic moments. This is the same mechanism that is behind the usual RKKY interaction, although basic features such as the characteristic $2k_F$ oscillations (with $k_F$ the Fermi momentum) will be absent in the spin liquid case and we can expect distinct consequences. In this work we shall explicitly calculate this spin-liquid-RKKY (slRKKY) interaction and investigate its impact on magnetic ordering of localised magnetic moments embedded in a spin liquid. With the aim of showing it is possible for such an interaction, under the right conditions, to drive nuclear magnetic order.

\begin{figure}
\begin{center}	\includegraphics[width=0.5\columnwidth]{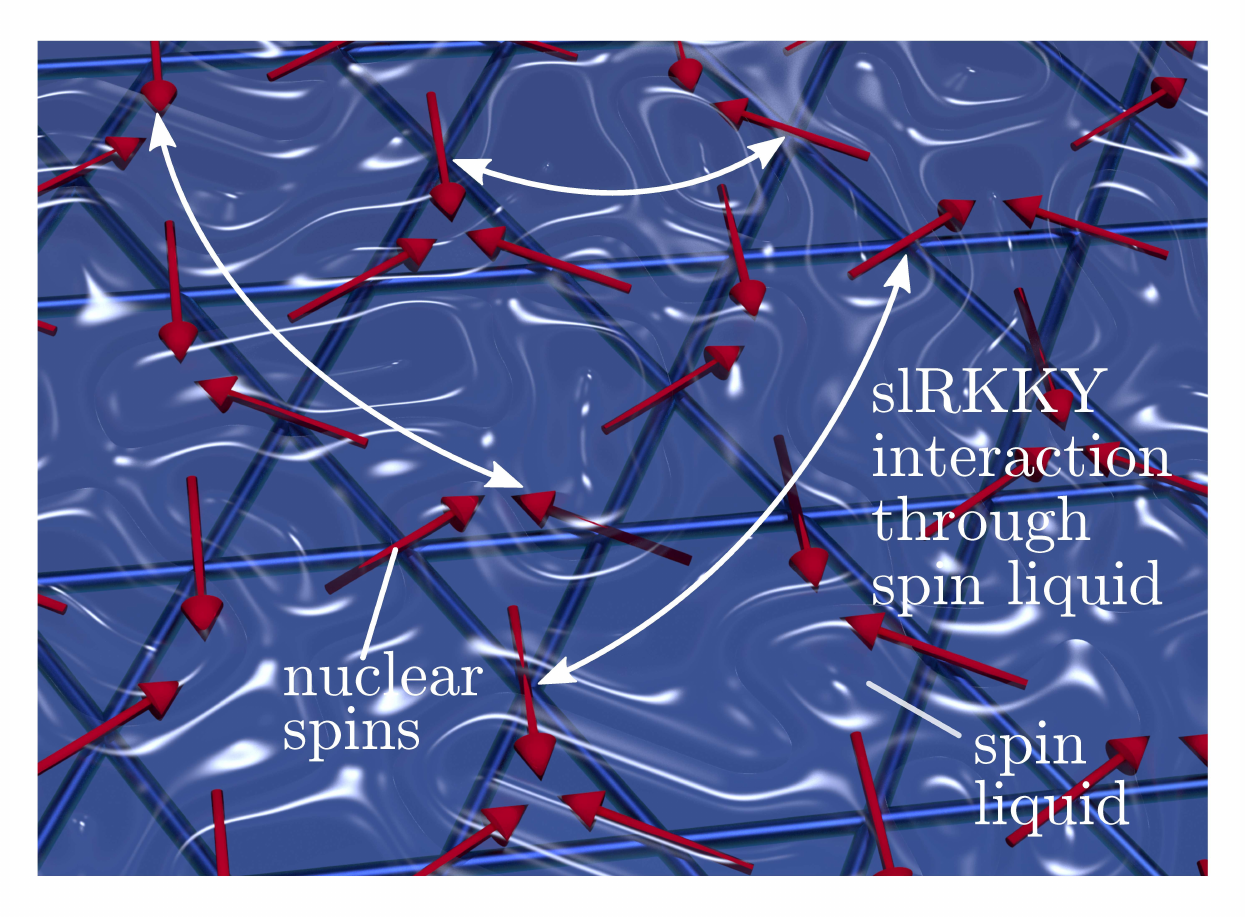}
	\caption{\label{fig:kagome}
	Nuclear spins (red arrows) embedded in a two-dimensional spin liquid
	(symbolised by the blue liquid) interact through slRKKY, an RKKY like effective interaction,
	(illustrated by the selection of white arrows) that is mediated through the correlated dynamics
	of the spin liquid. In the presence of a magnetic field this interaction can lead to a
	nuclear magnetic order as shown by the red arrows.}
\end{center}
\end{figure}

Our treatment will remain as general as possible and the consequences of such an interaction should also be broadly applicable to different types of localised moments and different spin liquids, of any dimension. Nonetheless, to have a concrete working example we will assume that the spin-half two dimensional kagome lattice holds a spin liquid and that this can couple to nuclear spins on the same sites as the electron spins, as shown in Fig. \ref{fig:kagome}. 

An alternative example in 2D would be the Kitaev model on the honeycomb lattice \cite{Kitaev:2006} and our analysis should also be extendible to three dimensional materials that may possess a quantum spin liquid, such as the hyper-kagome lattice \cite{finn:2016} or the 3D Kitaev materials.\cite{kevin:2016} However, the simplicity of kagome's geometric frustration and relatively simple lattice structure makes it a more tractable working example to analyse the slRKKY interaction in an explicit setting.

The analysis should also be applicable to different types of localised moments. Nuclear spins are, however, naturally embedded in the spin liquid on the ions of the crystalline lattice and so do not need to be added by further material engineering, as well as being accessible through the magnetic resonance techniques of NMR. The hyperfine interaction provides the coupling between the two spin species. Our analysis assumes from the start the existence of a quantum spin liquid and investigates what this spin liquid's consequences will be for the nuclear spins as a result of the associated slRKKY interaction.

In such a system the nuclear spins form a Kondo type lattice in which the spin liquid state takes the role of both the mediator of slRKKY interaction and of Kondo type screening. Although a variant of Kondo lattice physics or a variant of general Kondo physics in a spin liquid may be possible \cite{cass:1996,florens:2006,kolezhuk:2006, seok:2008,dhochak:2010,ribeiro2011,serbyn:2013,das2016,vojta:2016} our focus will be different because in most cases the nuclear spins are much larger than those that form the spin-liquid -- e.g. the $S=1/2$ of the kagome working example -- and as a result the semi-classical treatment used here becomes accurate. Additionally the hyperfine coupling between electron and nuclear spins is usually so small that the Kondo temperature -- which reduces exponentially with the inverse of the hyperfine coupling -- is much lower realistic temperature obtainable experimentally.

The results of this work are as follows: In the first half of the manuscript we show that an RKKY-type interaction can be mediated by the electron spins that form a spin liquid and we discuss the possibility for this interaction to induce nuclear magnetic order at low temperatures. In the second half we investigate a specific example, the KHAF, although our general treatment should remain applicable to other spin liquids. We demonstrate that for KHAF nuclear magnetic order is extremely fragile and the existence of a flat magnon band of spin-waves destabilises any nuclear order at any finite temperature, regardless of whether the system is infinite or finite. This fragility corroborates  the general phenomenon of the absence of order in low dimensions, but appears here for a long-ranged interaction in a non-itinerant system that is not covered by the Mermin-Wagner theorem\cite{mer:1966} or extensions.\cite{loss:2011} Despite this it is still possible for order to be stabilised at potentially experimentally obtainable temperatures, by applying a small magnetic field that gaps out these magnon fluctuations.

\section*{Order via slRKKY}
To begin we derive our main result: the general Hamiltonian for an slRKKY interaction between embedded nuclear spins and mediated by an underlying spin-liquid, such as the situation shown in Fig. \ref{fig:kagome}. We also discuss the possibility of nuclear magnetic ordering via such an interaction. Later in the manuscript we will then switch to a specific example, the kagome anti-ferromagnet, which we keep in mind during this section.
\subsection*{Spin liquid RKKY Hamiltonian}
 The Hamiltonian of the whole system in Fig. \ref{fig:kagome} -- spin liquid and its interaction with the nuclear spins -- can be written in the form
\begin{equation}
H=H_{\text{sl}}+H_{\text{hyp}},\label{Ham}
\end{equation}
where $H_{\text{sl}}$ is the Hamiltonian of the spin liquid and $H_{\text{hyp}}$ is the Hamiltonian of the hyperfine interaction between the electron and nuclear spins which we choose to be a contact interaction,
$H_{\text{hyp}}=A\sum_{i}{\bf S}_i \cdot {\bf I}_i$.
Here the strength of the the hyper-fine interaction $A$ is either positive or negative and ${\bf S}_i=(S^x_i,S^y_i,S^z_i)$ and ${\bf I}_i=(I^x_i,I^y_i,I^z_i)$ are the spin operators at lattice site $i$ of the localised electron and nuclear spins respectively. 

For simplicity we make two assumptions: 1) We consider nuclear spins located on the same site as the electron spin and assume that the nuclear moments are large and can be treated semi-classically. 2) We make the normal RKKY assumption that the hyperfine interactions is weak, in other words, $|A|/J \ll 1$, this is reasonable since in most spin liquid candidate materials $|A|/k_B$ is of the order of $10$mK whereas $J/k_B$ is of the order of $100$K (with $k_B$ the Boltzmann's constant). In this regime the electron spin relaxation times are much faster than typical time-scales of nuclear spin relaxation. 

Since the hyperfine coupling, $A$, is a small energy scale we can perform a Schrieffer-Wolff transformation, eliminating terms linear in $A$ and keeping terms up to order $A^2$. After integrating out the electron spin degrees of freedom we obtain an effective interaction Hamiltonian of the form 
\begin{equation}
H_{\text{SW}}=H_{sl}+\frac{1}{2}[K,[K,H_{\text{sl}}]],
\end{equation}
with $K$ defined by $H_{\text{hyp}}+[K,H_{\text{sl}}]=0$. This is solved such that
\begin{equation}
\frac{1}{2}[K,[K,H_{\text{sl}}]]=-\frac{i}{2}\int^{\infty}_0 dt e^{-\eta t} [H_{\text{hyp}}(t),H_{\text{hyp}}(0)],
\end{equation}
where the time evolution of the hyperfine Hamiltonian is given by $H_{\text{hyp}}=e^{i H_{\text{sl}}t} H_{\text{hyp}}(0) e^{-i H_{\text{sl}}t}$ and $\eta\rightarrow 0^+$ ensures convergence of the integral. Inserting $H_{\text{hyp}}$ gives
\begin{equation}
 \begin{aligned}\label{effective}
 \frac{1}{2}[K,[K,H_{\text{sl}}]]&=-\frac{i A^2}{2}\sum_{i,j}\int^{\infty}_0 dt\;e^{-\eta t}  [{\bf I}_i\cdot{\bf S}_i(t),{\bf I}_j\cdot{\bf S}_j(0)]\\
 &=-\frac{i A^2}{2}\sum_{i,j,\alpha,\beta}\int^{\infty}_0 dt\;e^{-\eta t} \bigg\{ I^{\alpha}_i I^{\beta}_j[S^{\alpha}_i (t), S^{\beta}_j]  + [I^{\alpha}_i, I^{\beta}_j] S^{\beta}_j S^{\alpha}_i(t) \bigg\},
 \end{aligned}
 \end{equation}
 where $\alpha,\beta=x,y,z$ are the spin components. The final step to obtain the effective slRKKY interaction is to integrate out the electron spin degrees of freedom from $H_0$, which leaves us with the real-space slRKKY Hamiltonian
 \begin{equation}
 \begin{aligned}
 H_{\text{eff}}&=-\frac{i A^2}{2}\sum_{i,j}\int^{\infty}_0 dt\; e^{-\eta t} I^{\alpha}_i I^{\beta}_j \langle[S^{\alpha}_i (t), S^{\beta}_j]\rangle_0\\
 &=\frac{A^2}{8J}\sum\limits_{i,j} \chi_{\alpha\beta}(i,j,\omega=0) I^{\alpha}_i I^{\beta}_j,\label{Hreal}
 \end{aligned}
 \end{equation}
 where  $\chi_{\alpha\beta}(i,j,0)$ is the real space static spin susceptibility and the second term of the last line of Eq. \eqref{effective} is zero due to the spin liquid being a non-magnetic phase. In the non-magnetic spin liquid state the susceptibility is independent of the directions in spin-space \cite{bernhard2002} so that $\langle S^+_i S^-_j\rangle=2\langle S^z_i S^z_j\rangle$, with $S^\pm_i=S^x_i\pm i S^y_i$. Hence we can write  $2\chi_{\alpha\beta}(i,j,0)=\chi_{+-}(i,j,0)=\chi(i,j)$, independent of the directions in spin space.

Since we are searching for long-ranged order the interaction of Eq.~\eqref{Hreal} is more easily dealt with in Fourier-space rather than real space.\cite{bernd:2009,bernd:2013,pasc:2008,loss:2011,simon:2007} This can be achieved by taking a Fourier transform of Eq.~\eqref{Hreal} with respect to the Bravais lattice upon which the electron and nuclear spin lattice is built
 \begin{equation}
{\bf I}^a_{\bf q}=\frac{1}{\sqrt{N_{Br}}}\sum\limits_{n} {\bf I}^a_n \; e^{-i {\bf q}\cdot {\bf R}_n } \qquad {\rm and} \qquad 
 \chi^{ab}({\bf q},\omega)=\frac{1}{N_{Br}}\sum\limits_{i,j} \chi^{ab}(i,j) e^{-i {\bf q}\cdot ({\bf R}^a_i -{\bf R}^b_j)},
\end{equation}
 where $N_{Br}$ is number of unit cells and $a,b$ refer to the lattice sites within the unit cell, for example in kagome they label the 3 sites of the unit cell within the dashed line indicated in Fig. \ref{fig:lattice}. The Fourier space slRKKY interaction is then
\begin{equation}
 H_{\text{eff}}=\frac{A^2}{4J} \sum\limits_{{\bf q},a,b}  \chi^{a b}({\bf q}) \; {\bf I}^{a}_{\bf q} \cdot {\bf I}^{b}_{\bf -q}. \label{Heff2}
\end{equation}
The highly correlated nature of the spin-liquid provides a susceptibility that is nonzero even at large length-scales and this enables the underlying spin-liquid to mediate a long-ranged interaction between local magnetic moments, as shown Eqs.~\eqref{Hreal} and \eqref{Heff2}. This interaction is the main result of this manuscript and we call such an interaction slRKKY. Moreover, since the mediator remains disordered at low temperatures, this long-ranged RKKY-like interaction provides the possibility of stabilising long-ranged nuclear magnetic order. 

\subsection*{Phenomenology of ordering via slRKKY}

The long-ranged slRKKY interaction of Eqs.~\eqref{Hreal} and \eqref{Heff2} has the potential to impose nuclear magnetic order. When considering the possibility of the appearance of long-range order in a low-dimensional system it is important to make the distinction between the possibility of order appearing in principle and the possibility of it appearing in practice.
Whether the appearance of order can occur in principle is addressed for a large class of systems by the Mermin-Wagner theorem.\cite{mer:1966} The theorem forbids long-range magnetic order in Heisenberg type systems when they satisfy the condition that the interaction is sufficiently short ranged (decaying faster than $1/r^{2+d}$ with $r$ the distance and $d$ the dimensionality of the system). The requirement for short range interactions has been partially lifted by an extension of the theorem by Loss, Pedrocchi, and Leggett,\cite{loss:2011} who rigorously proved the absence of long-range order in low dimensions for RKKY systems with arbitrary electron-electron interactions, provided that the interactions are isotropic and the RKKY interaction is carried by itinerant electrons.
Since the majority of spin liquids systems have localised electron spins this extended Mermin-Wagner theorem is not applicable here.

However, since the extended Mermin-Wagner theorem is invariant even when interactions are included it is likely that a further extension of the theorem will hold for slRKKY. Indeed within our example system, the KHAF, we show via explicit calculation that it is the case that any order breaks down if we maintain isotropy in nuclear spin space but that we can stabilise this order by breaking that isotropy with a small magnetic field.

Yet the exclusion of long-range order \emph{in principle} must further be contrasted with long-range order \emph{in practice}. If the order
destabilising fluctuations become effective only at wavelengths larger than the sample size then order can extend over the entire sample and remain
stable up to reasonable, accessible temperatures. For regular RKKY systems strong correlations induced by electron-electron interactions can play a
crucial stabilising role here, as was demonstrated for two-dimensional \cite{simon:2007,pasc:2008} and one-dimensional conductors.\cite{braunecker:2009a,bernd:2009,meng:2013,sch:2014} The generality of the ordering mechanism has been further exploited to find self-sustained 
topological phases when the electrons are driven into a superconducting state.\cite{bernd:2013,klin:2013,vaz:2013}
As most of this physics depends on the existence of an electronic Fermi surface, i.e. on $k_F$, a distinct phenomenology exists for slRKKY.

One notable consequence resulting from the lack of a Fermi-surface $k_F$ is that the minimum wavelength for destabilising fluctuations is set by the Brillouin zone boundaries and so even finite systems will have very low ordering temperatures unless they are extremely small - at which sizes they may no longer exhibit spin liquid behaviour. However the unique nature of slRKKY also enables a unique solution to the problem of ordering: By applying a magnetic field we can couple to the magnon spectrum of the nuclear spins and induce a gap, thereby ordering the nuclear spins below a temperature that scales with the nuclear Zeeman energy and is potentially accessible experimentally.

Another consequence of the nature of the spin-liquid mediator is that a nuclear spin order can exert a reciprocal action on the spin liquid, with the nuclear spins forming an effective magnetic field (Overhauser field). In principle this field could have a destabilising effect on the spin liquid. For our focus on the nuclear spins, such a back-action, however, is unlikely to affect the orderings of the nuclear moments considered here since the back action will be small due to $A \langle I \rangle \ll J$. On the other hand, this back-action may have consequences for the spin-liquid at temperatures well below the nuclear spin ordering temperature, but such a scenario is beyond the scope of this work and our focus will remain on the potential for nuclear spin order.

Whilst slRKKY has a distinct phenomenology in comparison to regular RKKY systems, its long-range also means it has a unique phenomenology when contrasted with lattice models containing only nearest-neighbour or short-ranged interactions. Unlike for short-ranged interactions, the natural choice is to consider the slRKKY in Fourier-space, this is because in principle a local excitation of the spin-liquid can couple to any lattice site of the underlying nuclear moments. A direct consequence of the non-locality of the effective interaction is that local deformations of spins are suppressed because ground states must satisfy energy constraints with the whole lattice, not just close neighbours. On the other hand such constraints are easily satisfied by periodic orderings, these are also the orderings most easily stabilised by anisotropies induced, for instance, by a magnetic field.

\section*{Example: The kagome antiferromagnet}

In this section we take an explicit example for our spin liquid Hamiltonian: the nearest-neighbour spin-1/2 antiferromagnetic Heisenberg model on the kagome lattice \cite{bernhard2002}
\begin{equation}
H_{\text{sl}}=\frac{1}{2}\sum\limits_{i,j}J_{ij}{\bf S}_i \cdot {\bf S}_j,\label{Heis} \qquad {\rm such\;that} \qquad J_{ij}=\begin{cases}
J>0, & \text{if $i$ and $j$ are nearest neighbors}\\
0, &\text{otherwise}.
\end{cases}
\end{equation}

\begin{figure}
\begin{center}
	\includegraphics[width=0.5\columnwidth]{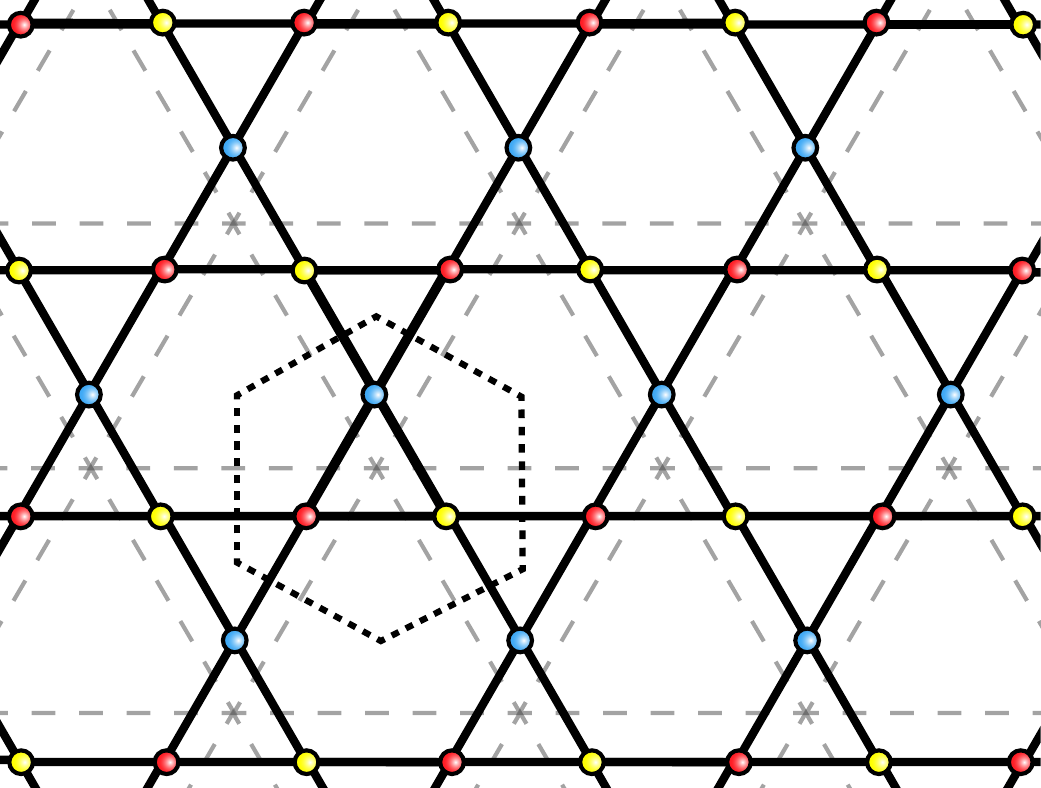}
	\caption{\label{fig:lattice}
	The kagome lattice, a tripartite lattice with three sites per unit cell -- here the unit cell is shown by the black dashed lines and sublattices by red, blue, and yellow points. It is built upon the triangular Bravais lattice -- here shown by gray dashed lines.}
\end{center}
\end{figure}

Kagome is made up of corner-sharing equilateral triangles with each site shared by only two elementary triangles (see Figs. \ref{fig:kagome} \& \ref{fig:lattice}). As such the kagome lattice takes advantage of the inherent geometric frustration of the triangular motif but has a reduced number of ground state constraints compared to the triangular lattice. This has the effect of increasing the size of the ground state degeneracy of kagome compared to that of the triangular lattice. It is for this reason that the $S=1/2$ nearest-neighbour antiferromagnetic Heisenberg model on the kagome  (KHAF) lattice is widely thought to contain a quantum spin liquid state at low temperatures, whereas the $S=1/2$ triangular lattice orders.\cite{sach:1992}

We should, however, stress that it has not yet been conclusively settled whether KHAF holds a spin liquid ground state or, if it does, the exact nature of this spin liquid.\cite{sach:1992,yan:2011,jia:2012,dep:2012,reuther:2014,mend:2016,sakai:2017, liao:2017} As such, our assumption for this as a working example should be understood in two ways: First, by assuming a spin liquid state, we can evaluate the consequences of the slRKKY interaction and this could be probed experimentally to check for consistency with our assumption. Second, we will see that the most important feature for the slRKKY interaction within our calculation will be the crystalline symmetries and the kagome lattice provides a non-trivial but still solvable illustration of this feature. 

Another motivation for using the KHAF as an example is that in the last decade several experimental KHAF candidates have been studied.\cite{mend:2016,zen:2001,faa:2012} Two notable examples are the `structurally perfect' \ce{Zn Cu_3 (OH)_6 CL_2} (Herbertsmithite) \cite{sho:2005} and \ce{[NH_4]_2 [C_7 H_{14} N][V_7 O_6 F_{18}]} (DQVOF).\cite{aid:2011} The suitability of these two candidates to hold a quantum spin liquid state is exemplified by the fact that no freezing of electron spins has been observed in Herbertsmithite or DQVOF down to temperatures of $50$mK and $40$mK respectively.\cite{mendels2007,clark2013}

\subsection*{slRKKY on KHAF}

Our derivation of Eqs.~\eqref{Hreal} and \eqref{Heff2} relied on two assumptions: 1) Large classical nuclear spins and 2) Weak hyperfine interactions. For the known KHAF candidates both are reasonable assumptions. For example in Herbertsmithite the equivalent nuclear moments to those discussed here correspond to $I=3/2$ copper ions and in DQVOF the $I=7/2$ vanadium ions.\cite{crc2014} Away from the KHAF, the $I=3/2$ iridium nuclear spins of the Kitaev iridates \cite{singh:2012} further show the more general applicability of our treatment to other potential spin liquid candidates. The second assumption is also reasonable for KHAF candidates, for example $J/k_B \sim 60$K is found in DQVOF.\cite{clark2013} 

\begin{figure}[h!]
\begin{center}
\includegraphics[width=0.5\columnwidth]{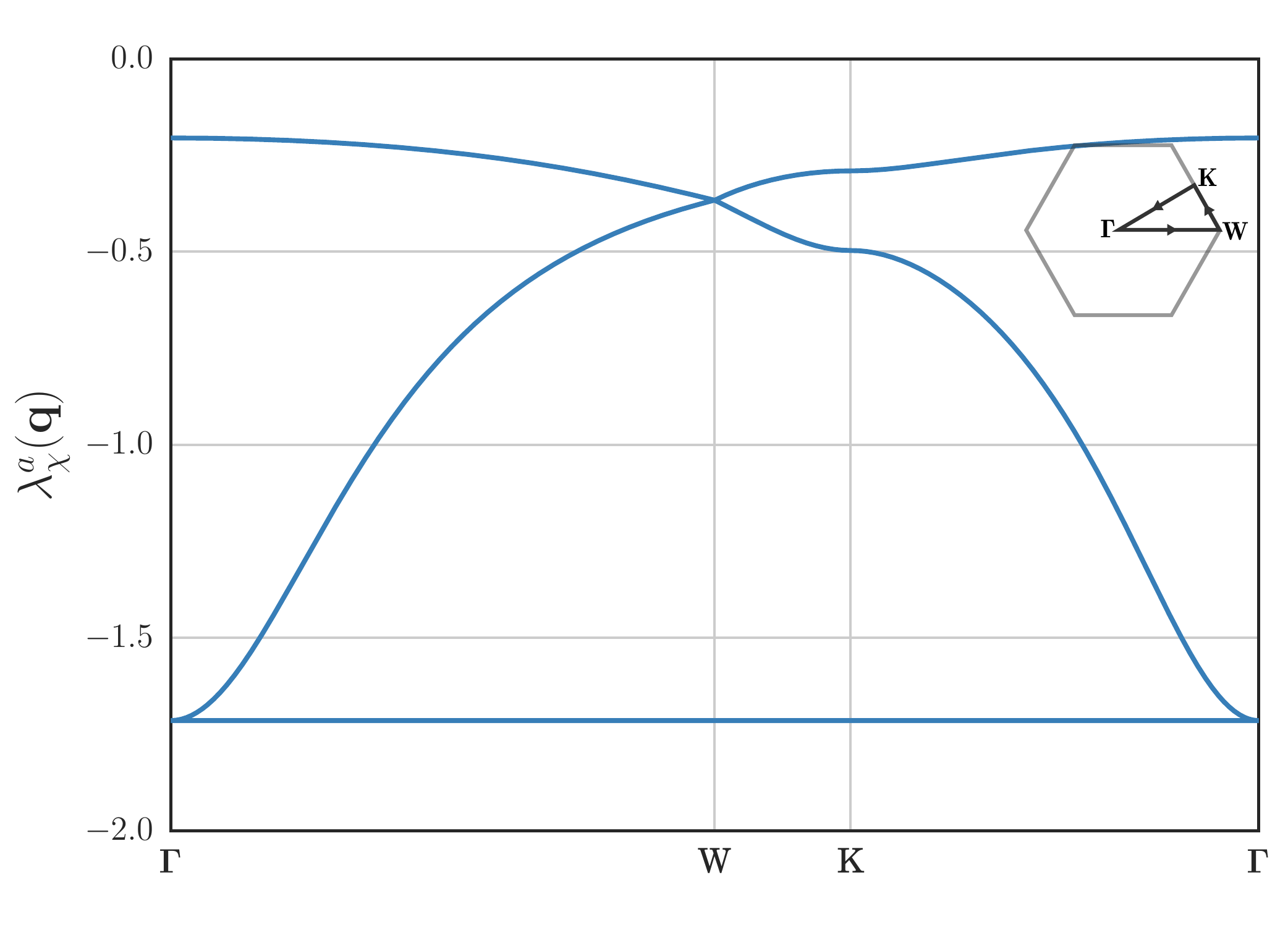}
\caption{The eigenvalues of the KHAF zero-temperature static susceptibility $\chi^{ab}({\bf q})$ along a high symmetry path in the Brillouin zone (shown inset), as calculated via a Kondo-Yamaji Green's function decoupling. The lowest eigenvalue is constant over the whole Brillouin zone.}
\label{fig:eigen}
\end{center}
\end{figure}

The only feature of the KHAF spin liquid entering the slRKKY Hamiltonian Eq. \eqref{Heff2} is the static susceptibility $\chi^{ab}({\bf q})$. As is the case for the standard RKKY,\cite{simon:2007} only the qualitative features of the susceptibility will enter our calculation, namely the positions of degenerate eigenvalues, in fact, unlike standard RKKY there is no $k_F$ and so the interaction is essentially entirely determined by symmetries of the susceptibility. In particular the positions of degenerate eigenvalues are set by the crystal symmetries of the lattice and, as a result, the orderings found in below are independent of the method used to calculate the susceptibility as long as it correctly captures the position of these degeneracies. For completeness in the supplementary material we present a  second-order Green's function decoupling method, first introduced by Kondo and Yamaji,\cite{tyablikov:1959,yamaji:1973} which can be used to calculate the key features of the spin susceptibility of the KHAF required for our analysis and obtain explicit analytical results from these features. This has previously been utilised to calculate the specific heat and structure factor of the KHAF.\cite{bernhard2002,yu2000} The eigenvalues of the susceptibility matrix  $\chi^{ab}({\bf q})$ as a function of indices $a$ and $b$ are shown in Fig.~\ref{fig:eigen}.

The key feature of the susceptibility shown in Fig.~\ref{fig:eigen} for our purposes is that it has an eigenvalue that is independent of ${\bf q}$. We can see from Fig. \ref{fig:eigen} that this is the lowest (largest in magnitude) eigenvalue of the susceptibility matrix.\cite{gar:1999} More accurate methods for calculating the spin susceptibility, for example a higher order decoupling or more modern RG methods,\cite{reuther:2014,finn:2016} may show that the flatness of the flat band is broken. Despite this, the Kondo-Yamaji decoupling provides a sufficient approximation to the true electron spin susceptibility for the purposes of calculating nuclear spin orderings resulting from slRKKY. This is because the positions of degenerate eigenvalues that are key to obtaining nuclear orderings are set by the kagome lattice symmetries and these are included in the Kondo-Yamaji scheme through the Heisenberg exchange matrices. As we show below, these symmetry arguments are applicable to the slRKKY interaction, Eq.~\eqref{Heff2}, because it is not frustrated in the same way as the NN KHAF. In particular slRKKY is long-ranged and so there is a coupling between all nuclear spins involving all bands of the spin susceptibility, not just a nearest neighbour coupling. If a different method where to find a deviation of the flat band this will either: 1) Have no effect on the nuclear spin orderings if the deviation is far from a degeneracy point or 2) if it is close to a degeneracy point, enable the slRKKY interaction energy to be minimised beyond our analysis in the following section and select an ordering (or reduced set of orderings) from the potential orderings we find.

\subsection*{Minimisation of slRKKY ground state energy}\label{LutTis}
To illustrate the impact of the slRKKY interaction we investigate the possibility of it inducing nuclear magnetic order.\cite{braunecker:2009a,simon:2007,pasc:2008,bernd:2009,meng:2013} This requires two steps: The identification of possible ground state orderings -- which is done in this section -- and the stability analysis of these orderings -- which is done in subsequent sections.

To minimise the ground state energy we find periodic spin configurations of lowest energy subject to the full classical constraint $|\vec I_i|^2 = I^2$ on each site (also called the ``hard constraint''). We must do this because the flat-band in the interaction means that the often used Luttinger-Tisza mean field method,\cite{luttinger:1946} which only includes a constraint on average spin-length (also called the ``weak constraint''), is not sufficient. Although in general a more complicated task, \cite{litvin1974} it is also the coincidence of the symmetries of the interaction and lattice that enables us to impose the full spin-length constraint and find periodic minimum energy states.

Due to its extended nature we expect that the ground states of slRKKY are periodic orderings since each configuration must minimise the energy on a global scale. This is because each site must satisfy energy constraints between itself and the whole lattice in contrast to, for example, the classical kagome antiferromagnet where line or local deformations of periodic states are possible due to its short range.\cite{chalk:1992} Without excluding the possibility of non-periodic states in principle, for answering the question of whether magnetic order is possible at all we shall focus on periodic ground states only. Therefore as the building blocks of slRKKY ground states we search for regular Bloch type periodic orderings of the form $\vec I^a_i=\vec u e^{-i \vec r_i\cdot \vec q}$. From these we construct planar spiral states $\vec I^a_i={\vec u}\cos({\vec q}\cdot( {\vec r}^a_i+{\vec T}^a))+{\vec v}\sin({\vec q}\cdot ( {\vec r}^a_i+{\vec T}^a))$ on each sublattice, where the $\vec T_a$ is the vector from the center of the unit cell to the lattice site, $\vec r_i$ the vector to the unit cell, $\vec q$ is a Bloch wave-vector (we will use $\vec Q$ for the wave vectors of specific orderings), and $\vec u$ and $\vec v$ orthogonal directions in spin space. In the supplement we will also show that a non-coplanar multiple-$\vec Q$ state can be constructed from these Bloch basis orderings. Later, however, we will show that the flat band in the interaction will ensure that any of the possible ground states will be unstable without the application of an additional perturbation.

In general periodic orderings are found by expanding the nuclear spin operators ${\bf I}^a({\bf q})$ on each sublattice $a$ in terms of the normalised eigenvectors $U_{a,\nu}({\bf q})$ of the susceptibility $\chi^{ab}({\bf q})$ where $a,b=1,2,3$ labels the sublattice component of the vector and $\nu=1,2,3$ labels the eigenvector. The expansion of the spins then reads,
$
{\bf I}_a({\bf q})=\sum_\nu{\bf W}_{\nu}({\bf q}) U_{a,\nu}({\bf q}),\label{eigenexp}
$
where the ${\bf W}_{\nu}({\bf q})$ are (orthogonal) directions in spin-space.

A lower bound for the ground state minimum energy is found by relaxing the (classical) constraint $|\vec I_i|^2=I^2$ to the constraint $\sum_i |{\bf I}_i|^2=N I^2$, so that only the average length of the spins in the entire system is equal to $I$.\cite{litvin1974} Inserting the expansion of the spins in terms of the eigenvectors of the susceptibility diagonalises the susceptibility matrix and we find a lower bound for the ground state energy $E_{\text{GS}}\geq A^2 I^2 N \lambda_{\text{min}} /4J\equiv E_{min},$
where $\lambda_{\text{min}}$ is the minimum eigenvalue of the susceptibility as shown in Fig. \eqref{fig:eigen}. For a standard Luttinger-Tisza analysis, based solely on this weaker constraint and in a Bravais lattice, this would specify ordering vectors $\{{\bf Q}_n\}$ where the minimum energy could be found. That is not the case here because the lowest eigenvalue of $\chi^{ab}({\bf q})$ is the same for all ${\bf q}$ (see Fig. \ref{fig:eigen}) and therefore there is no unique ordering vector or set of ordering vectors $\{{\bf Q}_n\}$ where this lower bound is achieved.

Instead we must search for the lowest energy configurations of nuclear spins by imposing the true constraint on the length of the nuclear spins at each site,\cite{maksy2015} $|{\bf I}_i|^2=I^2$, which in Fourier space reads
$
\frac{1}{N_{Br}} \sum_{{\bf k},{\bf G},a} {\bf I}^a_{\bf k}\cdot  {\bf I}^a_{\bf q-k}=\sum_{{\bf G}}I^2\delta_{\bf q G},\label{hconstraint}
$
where $a=1,2,3$ refers to the sublattice and ${\bf G}$ runs over reciprocal lattice vectors.

For a non-Bravais lattice, such as kagome, the full classical constraint on spin normalisation can only be fulfilled by a spin configuration that simultaneously achieves the minimum energy at a $\vec Q$ where the eigenvectors of the susceptibility matrix have the same weight on each site of the unit cell. If the eigenvectors had different weights on each site this would require the lengths of spins within a unit cell to vary such that they were inversely proportional to the magnitude of the corresponding component of the susceptibility eigenvector in order to achieve the minimum energy.\cite{litvin1974}

For the case of the slRKKY Hamiltonian in Eq. \eqref{Heff2}, where the lattice of the nuclear spins is the same as the lattice of the electron spins forming the spin liquid, this requires that minima are located at high-symmetry points in the Brillouin zone. The symmetry ensures the diagonals of the susceptibility satisfy the condition
$
\chi^{11}({\bf Q}_n)=\chi^{22}({\bf Q}_n)=\chi^{33}({\bf Q}_n),
$
 and the off-diagonals satisfy
$
 |\chi^{12}({\bf Q}_n)|=|\chi^{23}({\bf Q}_n)|=|\chi^{13}({\bf Q}_n)|,
$
where $n$ labels a high-symmetry point in reciprocal space.

The above constraints on the susceptibility matrix mean that the energy minima $E_{min}$ can only be achieved for the nuclear spin system at ordering vectors, ${\bf Q}_n$, where there is a degeneracy in the eigenvalues of the susceptibility matrix. From Fig. \ref{fig:eigen} we see that, within the Brillouin zone of the triangular Bravais lattice, this only occurs at the Brillouin zone center, the $\Gamma$ point, and the Brillouin zone corners, the $W$ points.

The flat band of the susceptibility leads to each individual high-symmetry point being a minimum of equal energy. Extra exchange interactions in a real material that are not nearest neighbour Heisenberg in nature and (weakly) break the flatness of this band can have the effect of selecting one of these minima by lowering it with respect to the others. As long as these extra exchange interactions are weak they will not move the minima from the high-symmetry points as these will remain the positions where a true spin expansion as in can be performed and hence the only positions where the lower bound for the ground state energy can be achieved. 

For our example we chose the nuclear spins so that they lie directly on the sites of the electron spins that form the kagome lattice. If the lattice of nuclear spins was not the same as that of the spin liquid, therefore holding a different set of symmetries to the slRKKY interaction, the presence of the flat band makes it likely that more complicated structures might be stabilised.\cite{dom:2008}

\subsection*{Orderings of the Nuclear Moments at Energy Minima}\label{orderings}
To obtain the periodic orderings that correspond to the energy minima at the high symmetry points we must analyse the eigenvectors of the susceptibility matrix at the ordering vector ${\bf Q}_n$, these eigenvectors tell us the relative Fourier-space angle between the three spins of the unit cell. As a result of considering individual sites of the unit cell the orderings must be calculated in the extended Brillouin zone. This extension is required because the Brillouin zone of the underlying triangular lattice does not include the complete information for structures which take into account individual sites of the kagome lattice.  

Fig. \ref{fig:eigen} shows that two separate scenarios can occur at the high symmetry points: either the lowest eigenvalue of susceptibility is not degenerate, as found at the corners of the first Brillouin zone, or it is two-fold degenerate, as found at the the $\Gamma$ point. This latter scenario actually enables the possibility of multiple-$\vec Q$ states (see supplement), however we are interested here only in regular periodic ordering states since these will be the easiest to stabilise against magnon fluctuations.

\begin{figure}[h!]
\begin{center}
\includegraphics[width=0.5\columnwidth]{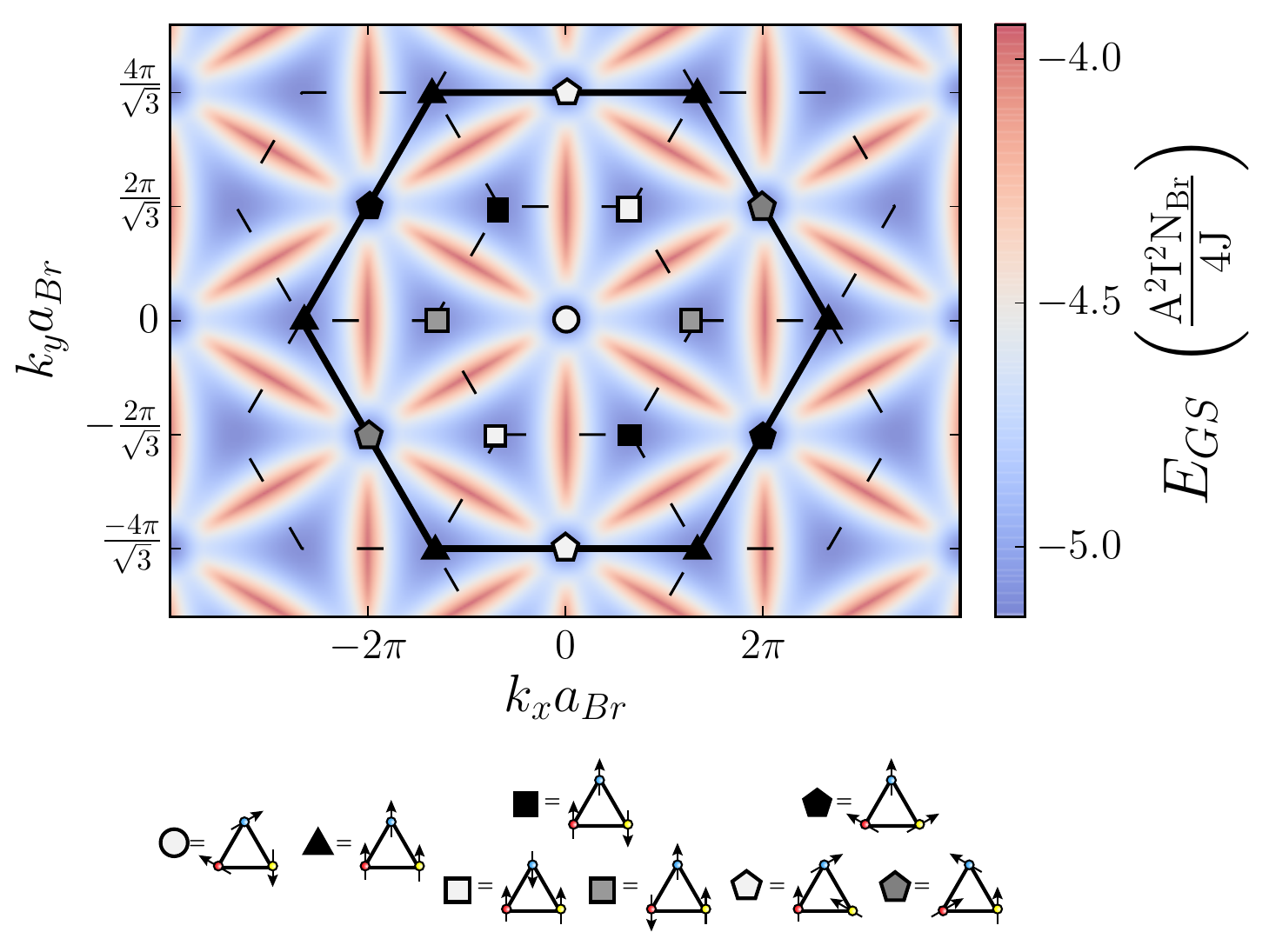}
\caption{Density plot of minimum energy in momentum space calculated by enforcing the hard spin length constraint for regular co-planar orders. The positions of ground state energies that achieve the lower bound of $E_{min}$ within the extended Brillouin zone (solid black line) are highlighted and we use these as the building blocks of slRKKY's ground states. The orderings, in the Fourier-space of the underlying Bravais lattice (before application of the phase that translates them to real space), at these points are shown in the key below.}
\label{fig:den}
\end{center}
\end{figure}

For the minima with a singular lowest eigenvalue the Fourier space expansion of the nuclear spin in terms of the eigenvectors of the susceptibility includes only one vector in spin-space. As such at non-degenerate high symmetry points all spins within the Fourier-space unit cell must be either parallel or anti-parallel. Only two momentum space orders are possible depending on the sign of susceptibility matrix elements at these points: either the spins create a two up--one down order with one spin in the unit cell anti-parallel to the other two, or they are all parallel. The former is found at the corners of the first Brillouin zone, whereas the latter is found at the corners of the extended Brillouin zone, these are shown in Fig. \ref{fig:den} as $\Box$s and $\triangle$s respectively.

For the high symmetry points where the lowest eigenvalue of the susceptibility is degenerate the expansion of the spins, two vectors in spin space are allowed. As such in Fourier space the nuclear spins within a unit cell can now lie at angles to each other in the same plane, with the angle given by the sign of the components of the eigenvectors of the susceptibility matrix. From the constraints on the elements of the susceptibility matrix we see the only possible momentum space orderings are all spins rotated $2\pi/3$ relative to each other, or a central spin with the other two spins rotated $\pm \pi/3$ to this central spin. The former is found at the $\Gamma$-point and the latter at the center of the edges of the extended Brillouin zone, these are shown in Fig. \ref{fig:den} as the $\Circle$ and $\pentagon$s respectively.

\begin{figure}[h!]
\begin{center}
\includegraphics[width=0.5\columnwidth]{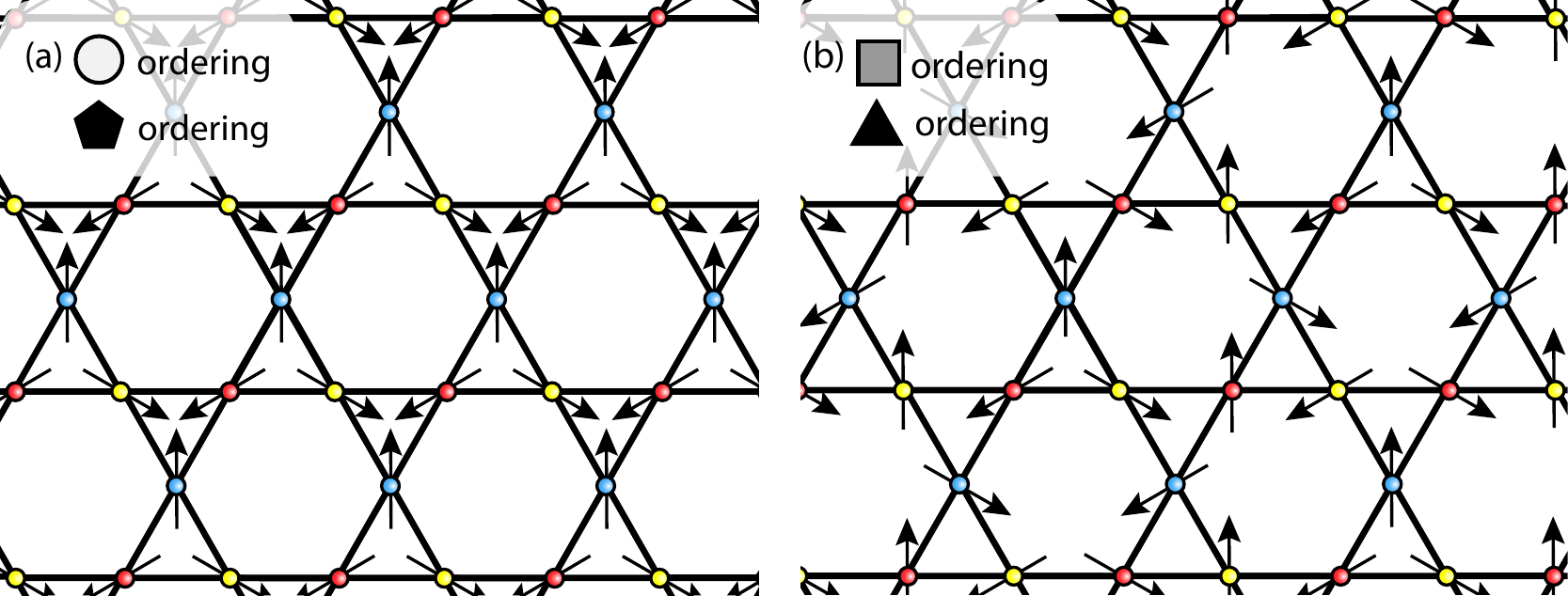}
\caption{The two regular periodic orderings of slRKKY, $\vec Q= 0$ and $\sqrt{3}\times \sqrt{3}$, shown in real space. The corresponding symbols from Fig. \ref{fig:den} are shown in the upper right corner of each pane.}
\label{fig:real}
\end{center}
\end{figure}

To translate these regular periodic orderings back into a real space ordering we place planar spirals on each sublattice $\vec I^a_i={\vec u}\cos({\vec Q}_n\cdot( {\vec r}^a_i+{\vec T}^a))+{\vec v}\sin({\vec Q}_n\cdot ( {\vec r}^a_i+{\vec T}^a))$ where the lattice vectors from the center of each unit cell to each site $\vec T^a$ are
\begin{equation}
{\vec T}^1 =\frac{a}{2\sqrt{3}}\hat{\vec y}, \;\; {\vec T}^2 =-\frac{a}{4}\left(\hat{\vec x}+\frac{1}{\sqrt{3}}\hat{\vec y}\right),\:{\rm and} \;\; {\vec T}^3 =\frac{a}{4}\left(\hat{\vec x}-\frac{1}{\sqrt{3}}\hat{\vec y}\right),
\end{equation}
with $a$ the lattice constant and $\hat{\vec x}$, $\hat{\vec y}$ the perpendicular unit vectors of the 2D plane.
This means that unit cells on planes separated by $\vec Q_n \cdot \vec r_m=2\pi m$, where $m$ is an integer, have each nuclear spin on equivalent sites aligned and the sites of the unit cells in any intermediate planes have their spins on each site rotated equally. Due to the phase $\vec Q_n \cdot {\vec T}^a$ we find that in fact in real space only two orderings are stabilised: $\vec Q=0$ order -- all spins on different sublattices at $120^\circ$ relative to each other and parallel within the sublattice -- and $\sqrt{3}\times \sqrt{3}$ order -- all spins in each plaquette at $120^\circ$ to each other with spins on the same sublattice rotated by $120^\circ$ between each site of the sublattice -- the former corresponds to the doubly degenerate $\Circle$ and $\pentagon$s and the latter the singular degenerate $\Box$s and $\triangle$s. Example orderings are shown in Fig. \ref{fig:real}. It should be noted that the spins are not constrained to the same plane as the kagome lattice because only the relative angle between spins is important to achieve an energy minimum. 


\subsection*{Stability of the nuclear moment orderings on kagome}\label{sec:SWs}
To test the stability of the mean field orderings found above against fluctuations, we investigate how spin-waves modify or destroy these orderings. To do this we select a basis for the directions in spin-space such that the spin on the first site of the Fourier-space unit cell is parallel to the spin z-axis, ${\bf I}_{1}({{\bf Q}_n})\parallel \hat{\bf z}$. Within this basis, because the regular periodic orderings are co-planar, we can effectively rotate the remaining spins in the unit cell about the y-axis to be parallel with the first spin by defining ${\bf I}'_{2}({{\bf Q}_n})=R(\theta_2){\bf I}_{2}({{\bf Q}_n})$ and  ${\bf I}'_{3}({{\bf Q}_n})=R(\theta_3){\bf I}_{3}({{\bf Q}_n})$, where $R(\theta_i)$ is the rotation matrix about spin y-axis and $\theta_i$ is the angle between the first spin and the spin on site $i$. Inserting this into Eq. \eqref{Heff2} gives,
\begin{equation}
\begin{aligned}\label{basisH}
H=\frac{A^2}{4J} \sum\limits_{{\bf q},a,b}  \chi^{ab}({\bf Q}_n)&\bigg\{\cos(\theta_a-\theta_b) \big(I'^z_a({{\bf Q}_n})I'^z_b(-{{\bf Q}_n})+I'^x_a({{\bf Q}_n})I'^x_b(-{{\bf Q}_n})\big)\\&+\sin(\theta_a-\theta_b) \big(I'^x_a({{\bf Q}_n})I'^z_b(-{{\bf Q}_n})-I'^z_a({{\bf Q}_n})I'^x_b(-{{\bf Q}_n})\big)+I'^y_a({{\bf Q}_n})I'^y_b({-{\bf Q}_n})\bigg\}.\\
\end{aligned}
\end{equation}
Since the nuclear spins are large we can take a $1/I$ Holstein-Primakoff expansion\cite{holstein1940} to estimate the nuclear spin-wave spectrum (see supplement). The zeroth order term simply gives the ground state energy $E_{min}$ and the linear term cancels, as expected for an expansion around a minimum. This leaves the second order terms of order $I$, which provide the excitation spectrum of the spin-waves, these can be written in as a block Bogoliubov matrix
\begin{equation}\label{SW-Ham}
H^{(2)}=\frac{A^2 N_{Br} I}{8J}\sum\limits_{{\bf q}}({\bf a}^{\dagger}_{-\bf q},{\bf a}_{\bf q})\left(\begin{array}{cc}
\boldsymbol{\Gamma}&\boldsymbol{\Lambda}\\ \boldsymbol{\Lambda}&\boldsymbol{\Gamma}
\end{array}\right)
\left(\begin{array}{c}
{\bf a}_{-\bf q}\\
{\bf a}^{\dagger}_{\bf q}
\end{array}\right).
\end{equation}
Here we defined on sublattice $a=1,2,3$ the vectors of bosonic operators 
$
\bigl[{\bf a}^{\dagger}_{\bf q}\bigr]^a=a^{a{\dagger}}_{\bf q}$,
 $\bigl[{\bf a}_{\bf q}\bigr]^a=a^{a}_{\bf q}$.
Similarly the $3\times 3$ matrices $\boldsymbol{\Gamma}$ and $\boldsymbol{\Lambda}$ have been defined with with entries
\begin{equation}\label{epsilon}
\left[\boldsymbol{\Gamma}\right]_{ab} =2 \delta_{ab}\sum\limits_c \cos(\theta_a-\theta_b)\chi_{ac}({\bf Q}_n) -\cos(\theta_a-\theta_b)\chi_{ab}({\bf Q}_n+{\bf q})
-\chi_{ab}({\bf Q}_n+{\bf q})\;\; {\rm and} \;\;
\left[\boldsymbol{\Lambda}\right]_{ab} =(-\cos(\theta_a-\theta_b)+1)\chi_{ab}({\bf Q}_n+{\bf q}).
\end{equation}

\begin{figure}[h!]
\begin{center}
         \includegraphics[width=1\textwidth]{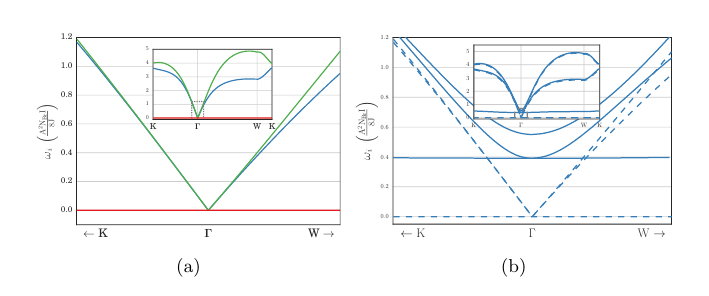}
     \caption{{\bf a)} The three spin-wave spectra for the ordering $\Circle$ centered about the the $\Gamma$-point. A zero-energy flat band is found at all ${\bf q}$ of reciprocal space. {\bf b)} The spin-wave spectra at finite field parallel to the central spin (blue) for the $\vec Q= 0$ ordering. The flat band of spin waves at no magnetic field is gapped out by a small finite field. (In both figures the full spectrum along the same high-symmetry path as Fig. \ref{fig:eigen} is shown inset.)}\label{fig:SW}
\end{center}
\end{figure}

The Hamiltonian is diagonalised via a Bogoliubov transformation for 3-species of bosons to find the spin-wave spectra $\omega_a ({\bf q})$, with $a=1,2,3$. The diagonalization is presented in the supplement. An example spectrum for the ${\bf Q}=0$ ordering is shown in Fig. \ref{fig:SW}a.

Remarkably we find that for each ordering there is a flat spin-wave band at strictly zero energy for all momenta ${\bf q}$ of reciprocal space. This zero-energy flat band in linear spin-wave theory means that at any finite temperature the occupation of magnon modes is extremely high and will quickly destabilise any order. In particular, for a system in the thermodynamic limit, the orderings found above will evaporate at any temperature since the thermal excitation of magnon modes will entirely destroy the order. This is a direct proof of the instability of long range nuclear magnetic order in this two-dimensional system. This provides a direct example where the Mermin-Wagner theorem can be extended to such slRKKY systems. A proof of the existence of a zero-energy eigenvalue can be found in the supplementary information.

The existence of the zero-energy band in the linear spin-wave theory is a direct consequence of the flat band of the kagome susceptibility. This is can be seen from the fact that only the hard constraint on spin length produced the energy minimum and such a constraint can only be enforced by terms beyond linear spin-waves, which we prove in the supplement. In other words the fluctuations to linear order reduce the length of spin in such a way that it is unaware of the hard constraint on spin length and because this is the origin of the ordering minimum it is therefore unsurprising that to linear order we find a flat zero-energy band. This not only holds for the regular periodic orderings but also for any multiple-$\vec Q$ state since all orderings are only enforced by the hard constraint on spin-length (see supplement).

If higher orders of the spin-wave expansion or exchange interactions that are not nearest neighbour Heisenberg or other corrections to the susceptibility were included, these could break the flatness of this spin-wave band, however the curvature will remain very small and will be on the size of the Brillouin zone. The corresponding ordering temperature in the thermodynamic limit will remain zero because a $\sim q^2$ dispersion still leads to a divergence in magnon occupation for 2D systems.\cite{simon:2007,pasc:2008} In a finite system the size of the system cuts off the longest possible wave length of fluctuations meaning that the temperature could be finite but will be extremely low due to the small amount of curvature in the band, even after other interactions or higher order spin-wave theory terms are included.

It is however possible to aid the slRKKY interaction in ordering nuclear spins by use of a magnetic field, which stabilises some of the ground state orders found in Fig. \ref{fig:den}. The magnetic field does this by creating a gap in the associated spin wave spectrum and therefore reducing the effects of thermally activated spin-waves on the order, meaning they do not evaporate as temperature is increased. It should be noted that although a small change in the ground state energy occurs, the main effect of the field is to stiffen the spins against fluctuations. In particular, the field does not significantly change the order itself since any nuclear Zeeman energy is smaller than the ground state energy of the order and so the effects of the slRKKY are remain dominant. Hence the main consequence of the magnetic field presented here is to break rotational symmetry in spin space. Such an effect could also naturally occur within real materials due to an exchange anisotropy on top of $J$ such as those that occur in Herbertsmithite,\cite{rig:2007,zor:2008, han:2012} although we do not intend to enter a material specific discussion and so the magnetic field presented here is therefore the most generic way of achieving such an effect. Of course the magnetic field (or exchange anisotropy) will also affect the KHAF spin liquid, however it has been shown that the small magnetic fields required here do not cause the breakdown of the spin liquid on the KHAF.\cite{nish:2013}

A uniform magnetic field across the whole lattice induces a nuclear Zeeman term $-\sum_a B I^z_a({\vec Q=0})$. The spin waves of all states are affected by the field, but since the magnetic field acts globally ($\vec Q=0$ in Fourier-space) the largest gap occurs for the $\vec Q=0$ state and so this periodic state will have the highest ordering temperature.

The second order terms of the Holstein-Primakoff expansion, which determine the magnon spectrum, will now be of the form (see supplement for details)
\begin{equation}
H^{(2)}=H^{(2)}(B=0)-g_{\text{nuc}}BN_{Br}I\sum\limits_{a,{\bf q}} a^{\dagger a}_{\bf q}a^{a}_{\bf q},
\end{equation}
where $H^{(2)}(B=0)$ is the zero-field RKKY term at quadratic  order. The result is that a constant is added to the terms diagonal in spin-wave operators and this opens a gap of the order of the order of $BN_{Br}I$ to the the spin-wave spectra -- including the flat band which previously destabilised the ordering at zero-field. As a consequence the ordering is stable until the temperature is of the order of the gap, at which point the lowest band in the spin-wave spectra is quickly occupied and the order will be destroyed. 

Hence we have shown that it is possible, under the right conditions, to stabilise nuclear magnetic order which is imposed via an slRKKY interaction. Using DQVOF as an example, the nuclear g-factor of vanadium \cite{crc2014} is $g_{nuc}=1.47$
and so would require a field $B\sim 500 \text{mT}$ per mK. Therefore, for small fields nuclear order via slRKKY could be stabilised at what correspond loosely to the limit of current state of the art experiments.
 \section*{Discussion}
We have shown that a spin liquid can mediate an RKKY-type interaction between localised moments such as nuclear spins. This interaction has a distinct character to the standard RKKY interaction of fermionic systems; for example, the absence of a $k_F$  means that the minimum wavelength for destabilising fluctuations is set by the Brillouin zone boundaries.

We choose the 2D kagome antiferromagnet as a working example to demonstrate the consequences of such an slRKKY interaction in an archetypal system that is thought to hold a spin liquid ground state. Although the existence of such a ground state has not been conclusively proven for the KHAF, its simple geometric frustration makes it a very tractable example to analyse what the influence of such a spin liquid would have on underlying nuclear spins. Many of the features of the electron spin susceptibility, which governs the strength of the slRKKY interaction, result directly from lattice symmetries. As a result, without full knowledge of the exact susceptibility of the true KHAF ground state, the nuclear spin orderings resulting from the slRKKY interaction can be calculated.

This KHAF example illustrates that such an interaction can induce distinct and varied orderings of nuclear moments. Interestingly, the lattice symmetries also cause a zero-energy flat magnon band, these orderings destabilise at any finite temperature regardless of if the system is infinite or finite in size. Despite this we also show that a small magnetic field can stabilise such an ordering to within potentially experimentally achievable temperatures without altering the underlying physics. This means, as a result of recent advances in developing spin liquid materials, such an interaction could have directly testable experimental consequences with today's state of the art.

Our treatment throughout this work has been general and we expect such an analysis should also be extendible and applicable to other spin liquids such as the 3D hyperkagome lattice and the Kitaev model realised on 2D and 3D lattices.

{\bf Acknowledgements}: The authors would like to thank A. Rosch, B. Normand, C.A. Hooley, and K. O'Brien for helpful discussions. H.F.L. thanks the DFG under CRC 1238 and Bonn-Cologne Graduate School of Physics and Astronomy (BCGS) for their financial support.

{\bf Author Contributions:} B.B: Conceived the project. H.F.L. and B.B.: Developed the theory. H.F.L. and B.B.: Wrote the manuscript. 

{\bf Competing Interests:} The authors declare no competing interests.

\bibliography{kagome-paper} 
\end{document}